# Implementation of the Orthodoxy Test as a Validity Check on Experimental Field Emission Data

Mohammad M. Allaham[1], Richard G. Forbes[2], Alexandr Knápek[3], Marwan S. Mousa[1*]

[1]*Surface Physics and Materials Technology lab, Department of Physics, Mutah University, Al-Karak 61710, Jordan.*

[2]*Advanced Technology Institute & Department of Electrical and Electronic Eng., Faculty of Eng. & Physical Sciences, University of Surrey, Guildford, Surrey GU2 7XH, UK.*

[3]*Institute of Scientific Instruments of the CAS, Královopolská 147, 612 64 Brno, Czech Republic*

*mmousa@mutah.edu.jo

**Abstract:**

In field electron emission (FE) studies, it is important to check and analyse the quality and validity of experimental current-voltage data, which is usually plotted in one of a small number of standard forms. These include the so-called Fowler-Nordheim (FN), Millikan-Lauritsen (ML) and Murphy-Good (MG) plots. The Field Emission Orthodoxy Test is a simple quantitative test that aims to check for the reasonableness of the values of the parameter "scaled field" that can be extracted from these plots. This is done in order to establish whether characterization parameters extracted from the plot will be reliable or, alternative, likely to be spurious. This paper summarises the theory behind the orthodoxy test, for each of the plot forms, and confirms that it is easy to apply it to the newly developed MG plot. A simple web tool has been developed that extracts scaled-field values from any of these three plot forms, and tests for lack of field emission orthodoxy.

## 1. Introduction

This paper discusses a simple new methodology for processing *measured* current-voltage $I(V)$ data from devices or systems that involve the process of *field electron emission* (FE) [1-5]. Note that in this paper the symbols $I$ and $V$ always denote the measured quantities that in some recent papers (e.g. [6]) have been denoted by $I_m$ and $V_m$. The symbols $I$ and $V$ do **not** denote the so-called "emission" quantities $I_e$ and $V_e$.

In FE literature, two types of plot have been used to analyse $I(V)$ data, namely Fowler-Nordheim (FN) plots [6-8] and Millikan-Lauritsen (ML) plots [9,10]. A third plot form, the Murphy-Good (MG) plot, has recently been proposed [11]. If the FE device/system is *orthodox*, as defined below, then all these plots present the $I(V)$ data as a nearly straight line that can be subjected to mathematical analysis, in order to extract emitter characterization parameters.

An FE device/system is defined as "ideal" if its $I(V)$ characteristics are determined only by the combination of: (a) unchanging total system geometry (including emitter shape); (b) unchanging emitter surface composition; and (c) the related electron emission process. The emitting system is further described as "orthodox" if it is adequately valid to assume that tunnelling takes place through a Schottky-Nordheim (SN) ("planar image-rounded") barrier, and that there is no significant voltage dependence in the emission area or in the local work-function. If a device/system is not orthodox, then data plots as discussed above may be defective, and extracted characterization-parameter values may be spurious.

There exists an "Orthodoxy Test" [12], developed in the context of FN and ML plots, that can be applied to an experimental FE $I(V)$ data-set, in order to establish whether or not the related FE device/system is orthodox, and hence whether extracted characterization-parameter values would be reliable. For example, there is some evidence [12] that many published field-enhancement-factor values may be spuriously large.

The present work discusses a simple web tool that can apply the orthodoxy test to any of the above plots, including the new MG plot. Relevant theory has been discussed elsewhere [8,12,13] and is summarised below. The orthodoxy test and web tool are then described and applied to illustrative examples of $I(V)$ data.

A motivation for this work has been to enhance the procedures available for testing field electron sources under development for possible use in electron microscopes and other electron beam instruments.

This paper uses the common "electron emission convention", whereby fields, currents, and current densities are treated as positive, even though they would be negative in classical electromagnetism. Where values of universal constants are given, numerical values are specified to seven significant figures.

## 2. Theory of extracting scaled-field values

### 2.1 Basic field electron emission theory

For an orthodox FE device/system, the measured current *I* can be given, using the local work function $\phi$ and the characteristic local barrier field $F_C$, by the *Extended Murphy-Good (EMG) equation* [11, eq. (2.2)]. It is better here to employ a "scaled" form that uses the dimensionless *characteristic scaled field* $f_C \equiv c_S^2 \phi^{-2} F_C$, where $c_S$ is the *Schottky constant* (see Appendix A). For orthodox systems, $f_C$ can be related to the measured voltage *V* by

$$f_C = V/V_R, \tag{1}$$

where $V_R$ is a (constant) *reference measured voltage* [11] needed to pull the top of a characteristic SN barrier, of zero-field height $\phi$, down to the emitter Fermi level. The EMG equation for $I(V)$ can thus be written [11]

$$I(V) = A\,\theta \cdot (V/V_R)^2 \exp[-v_F \eta V_R/V], \tag{2}$$

where *A* is a parameter called the *formal emission area for the SN barrier* (denoted by $A_f^{SN}$ in [11]), $\theta$ and $\eta$ are $\phi$-dependent scaling parameters defined in [11], with $\eta \cong 9.836239\,(\mathrm{eV}/\phi)^{1/2}$, and $v_F$ is the appropriate value of a special mathematical function $v(x)$ [14]. It can be shown that $v_F = v(x = f_C)$.

A simple good approximation exists [15] namely: $v_F \approx 1 - f_C + (f_C/6)\ln(f_C)$. Substituting into (2), and using (1) again, yields, after some re-arrangement, the alternative format

$$I(V) = \{A\,(\theta \exp\eta) V_R^{-\kappa}\}\, V^\kappa \exp[-\eta V_R/V], \tag{3}$$

where, for this SN-barrier case, $\kappa = 2 - \eta/6$.

In general terms, what the orthodoxy test does is to deduce, using the slope of a given plot form, the range of values of $f_C$-values that corresponds to the range of measured-voltage used in the experiments. As discussed below, this extracted range of $f_C$-values is then compared with the known range of $f_C$-values within which (tungsten) emitters normally operate.

Theory relating to the plot slopes is now given. In what follows, subscripts such as "FN" label the plot type, and the notation $\ln\{Q\}$ means (see [16]) "take the natural logarithm of the numerical value of the quantity $Q$, when this quantity is measured in SI units" (here amperes, and volts to an appropriate power).

**2.2 The theoretical Fowler-Nordheim (FN) plot slope**

With (2), on dividing both sides by $V^2$ and taking natural logarithms, we obtain

$$L_{FN}(V^{-1}) \equiv \ln\{I/V^2\} = \ln\{A\theta V_R^{-2}\} - v_F \eta V_R V^{-1}.$$

This is a theoretical equation for a Fowler-Nordheim (FN) plot. Its slope is given by

$$S_{FN}(V^{-1}) = dL_{FN}/d(V^{-1}) = -\eta V_R d(v_F V^{-1})/d(V^{-1}).$$

A standard result [15] is that $d(v_F V^{-1})/d(V^{-1}) = s(f_C)$, where $s(f_C)$ is the slope correction function for a SN barrier, and $f_C$ corresponds to $V$. Hence, the FN-plot slope is

$$S_{FN}(V^{-1}) = -s(f_C) \cdot \eta V_R. \tag{4}$$

In FN-plot analysis, the slope is (in principle) taken at the "fitting value" where the tangent to the theoretical plot is parallel to the line fitted to the experimental results [15]. This fitted line has slope $S_{FN}^{fit}$, and the fitting value of $s(f_C)$ is denoted by $s_t$. It follows from (4) that the extracted $V_R$-value is $\{V_R\}^{extr} = -S_{FN}^{fit}/s_t\eta$, and hence the extracted $f_C$-value corresponding to measured voltage $V$ is

$$\{f_C\}^{extr} = -(s_t\eta/S_{FN}^{fit}) V = -(s_t\eta/S_{FN}^{fit})/V^{-1}. \tag{5}$$

Since $s(f_C)$ varies only weakly with $f_C$, it is normally adequate to take $s_t = 0.95$.

### 2. 3 The theoretical Millikan-Lauritsen (ML) plot slope

Using (2), and taking natural logarithms of both sides, yields

$$L_{\mathrm{ML}}(V^{-1}) \equiv \ln\{I\} = \ln\{A\theta V_R^{-2}\} - 2\ln\{V^{-1}\} - v_F \eta V_R V^{-1}].$$

This is the theoretical equation for a Millikan-Lauritsen (ML) plot. Its slope is

$$S_{\mathrm{ML}}(V^{-1}) = dL_{\mathrm{ML}}/d(V^{-1}) = -2V - \eta V_R d(v_F V^{-1})/d(V^{-1}) = S_{\mathrm{FN}}(V^{-1}) - 2V.$$

Let $V_{\mathrm{mid}}^{-1}$ be the midpoint of the range of values of $V^{-1}$ used in an experimental ML plot. It follows that the slope $S_{\mathrm{FN}}^{\mathrm{eff}}$ of the corresponding FN plot is given approximately by

$$S_{\mathrm{FN}}^{\mathrm{eff}} \approx S_{\mathrm{ML}}^{\mathrm{fit}} + 2V_{\mathrm{mid}}.$$

Values of scaled field can be extracted by using (5), with $S_{\mathrm{FN}}^{\mathrm{fit}}$ replaced by $S_{\mathrm{FN}}^{\mathrm{eff}}$.

### 2.4 The theoretical Murphy-Good (MG) plot slope

Dividing both sides of (3) by $V^\kappa$, and taking natural logarithms, yields

$$L_{\mathrm{MG}}(V^{-1}) \equiv \ln\{I/V^\kappa\} \approx \ln\{A \cdot (\theta \exp\eta) V_R^{-\kappa}\} - \eta V_R V^{-1}.$$

This is the equation for a theoretical Murphy-Good (MG) plot. Its slope is [11]

$$S_{\mathrm{MG}}(V^{-1}) = -\eta V_R.$$

Using (1), the extracted $f_C$-value corresponding to a given $V$-value is

$$\{f_C\}^{\mathrm{extr}} = -(\eta/S_{\mathrm{MG}}^{\mathrm{fit}})/V^{-1}. \tag{6}$$

As compared with (5), the factor $s_t$ is not present.

For the extraction of other characteristic parameters see [6, 8].

## 3. Applying Test Criteria

For new data plots, use of the Murphy-Good (MG) plot [11] is recommended, because this method of extracting formal emission areas is more precise. However, all three experimental plot types will be approximately straight for an orthodox emitter. A straight line can be fitted either manually (this is usually good enough), or by a regression calculation.

To apply the orthodoxy test, the web tool will first calculate the slope of the fitted line from entered values of the plot's upper left ("up") and lower right ("low") ends. Figure 1 shows an example [13] of a MG plot with the required points to apply the test marked.

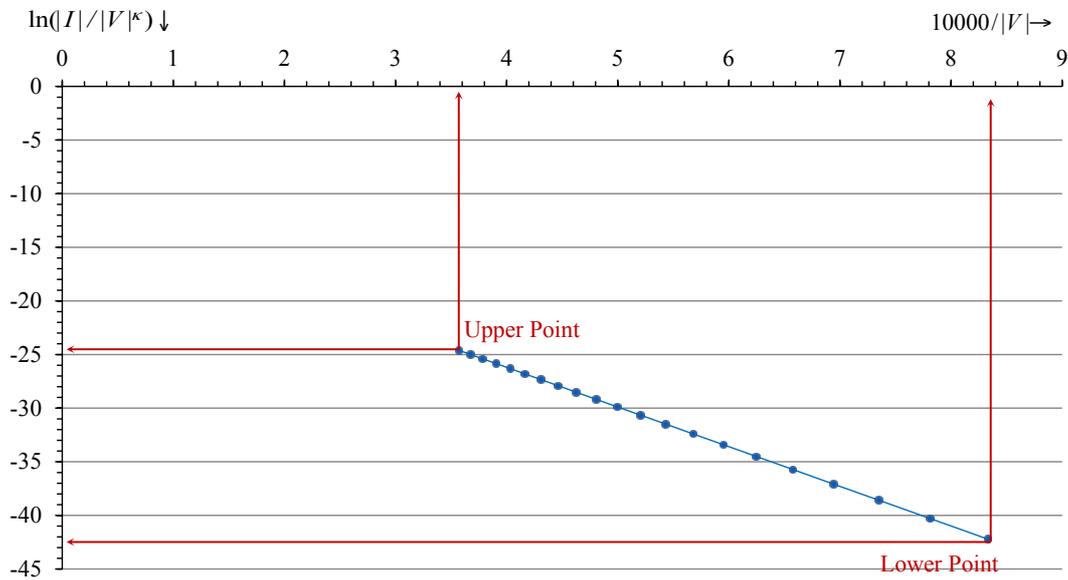

Fig. 1.   Simulated MG plot showing the upper-left and lower-right data points that need to be extracted and entered. The symbols $|V|$ and $|I|$ denote the numerical values of measured voltage and current when recorded in volts and amperes, respectively. The plotted points are selected points in the voltage range 1.2 to 2.8 kV.

After calculating the plot slope, the $f_C$-values corresponding to the ends of the range of voltages measured are extracted, using (5) or (6) as appropriate, depending on the plot type. The web tool will then apply the test criteria shown in Tab. 1 [12, 13]. Here: A/NA indicates the allowed/not-allowed limits for $f_C$; the parameter $\{f_{\text{low}}\}^{\text{extr}}$ is the extracted $f_C$ value for the lower-right point, and $\{f_{\text{up}}\}^{\text{extr}}$ is the extracted $f_C$ value for the upper-left point. Table 2 shows how the A/NA limits vary as a function of local work function $\phi$ (interpolation can be used if needed).

Table 1. General Criteria for the orthodoxy test.

| Condition | Result | Explanation |
|---|---|---|
| $f_{\text{low}}^{\text{A}} \leq f_{\text{low}}^{\text{extr}}$ AND $f_{\text{up}}^{\text{extr}} \leq f_{\text{up}}^{\text{A}}$ | Pass | Reasonable range |
| $f_{\text{low}}^{\text{extr}} \leq f_{\text{low}}^{\text{NA}}$ OR $f_{\text{up}}^{\text{NA}} \leq f_{\text{up}}^{\text{extr}}$ | Fail | Clearly unreasonable range |
| $f_{\text{low}}^{\text{NA}} \leq f_{\text{low}}^{\text{extr}} \leq f_{\text{low}}^{\text{A}}$ | Inconclusive | More investigation is needed |
| $f_{\text{up}}^{\text{A}} \leq f_{\text{up}}^{\text{extr}} \leq f_{\text{up}}^{\text{NA}}$ | Inconclusive | More investigation is needed |

Table 2. Range limits for the orthodoxy test, as a function of work function $\phi$. (Symbol meanings as defined in the text.)

| $\phi$ (eV) | $f_{\text{low}}^{\text{NA}}$ | $f_{\text{low}}^{\text{A}}$ | $f_{\text{up}}^{\text{A}}$ | $f_{\text{up}}^{\text{NA}}$ |
|---|---|---|---|---|
| 5.50 | 0.09 | 0.14 | 0.41 | 0.69 |
| 5.00 | 0.095 | 0.14 | 0.43 | 0.71 |
| 4.50 | 0.10 | 0.15 | 0.45 | 0.75 |
| 4.00 | 0.105 | 0.16 | 0.48 | 0.79 |
| 3.50 | 0.11 | 0.17 | 0.51 | 0.85 |
| 3.00 | 0.12 | 0.18 | 0.54 | 0.91 |
| 2.50 | 0.13 | 0.20 | 0.59 | 0.98 |

The physical meanings of the "not allowed" limits for the lower and upper points are as follows. The lower limit corresponds to the value where the field is too low to emit a current that can be measured or detected in a normal experiment. The upper limit corresponds to the value where the emitter will electroform or self-destruct. In both cases, if any extracted $f_C$-value is on the "not-allowed" side of the limit, then it can be concluded that the FE device/system is not orthodox, and that extracted values of emitter characterisation parameters may be spurious [12].

**4. Results and Examples**

During the project reported in this paper, the orthodoxy test was applied to many experimental and simulated data plots, using the web tool [17] in its state as developed. Data relating to the orthodoxy test is displaying correctly. Our plan is to extend this tool, in order to extract characterisation parameters from plots that pass the orthodoxy text, but this aspect of the tool is still under development and (at the time of writing) related "boxes" may either be blank or may not be displaying meaningful data.

### 4.1 Murphy-Good (MG) Plot Analysis

Figure 1 above shows a simulated MG plot. Extracted input data for the orthodoxy test, and output results, as associated with the web tool [17], are recorded in Tabs 3 and 4. As expected with simulated data, the result is "PASS".

### 4.2 Millikan-Lauritsen (ML) Plot Analysis

The spreadsheet originally developed in connection with the orthodoxy test (see Electronic Supplementary Material to [12]) has been extensively used to test ML plots. We therefore confirm here, for one example only, that the new web tool generates the same result as the original 2013 spreadsheet. The chosen example is emitter X89 (Fig. 4) in the well-known paper by Dyke and Trolan [18]. The relevant data (relating to the "direct-current" voltage range) are re-entered in Tabs 3 and 4. It has been confirmed that the present web tool gives the same extracted $f_C$-values as the original spreadsheet.

### 4.3. Fowler-Nordheim (FN) Plot Analysis

With FN-plot analysis we show an example of an "inconclusive" outcome. Curve A in Fig. 2 shows a FN plot for an "uncoated" tungsten emitter. Relevant data are shown in Tables 3 and 4. Although the FN plot is a good straight line, the orthodoxy test reports that the outcome is INCONCLUSIVE. The most likely explanation is that the emitter was being operated up to higher fields than is usually the case—possibly up to higher fields than would usually be advisable.

Many examples of FN plots that fail the orthodoxy test are given in [12].

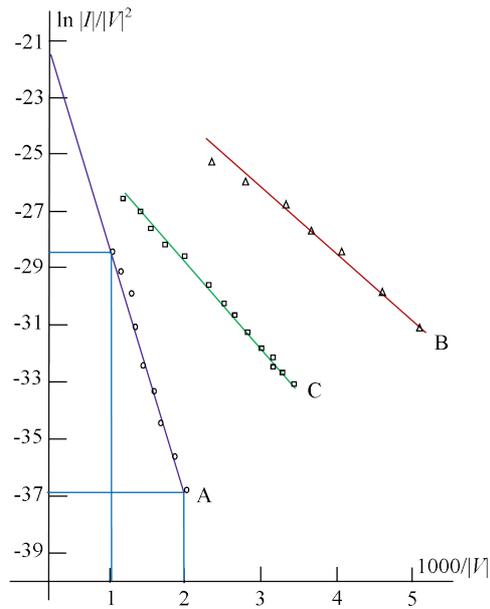

**Fig. 2.** Curve A shows a Fowler-Nordheim plot taken from an uncoated tungsten emitter [19]. The symbols |*V*| and |*I*| denote the numerical values of measured voltage and current when recorded in volts and amperes, respectively.

### 4.4 Outcomes and Discussion

For convenience, the outcomes from the above illustrative tests are shown together in Tables 3 and 4 below.

**Table 3**. Input data for orthodoxy test. Note that the neper (Np) is a unit of natural logarithmic difference (here relative to *Y*=0)

| Plot Type | Source | Figure here | Upper (Left) Point $V^{-1}$ (V$^{-1}$) | $Y$ (Np) | Lower (Right) Point $V^{-1}$ (V$^{-1}$) | $Y$ (Np) |
|---|---|---|---|---|---|---|
| MG | Simulated | Fig. 1 | 3.57×10$^{-4}$ | −24.6 | 8.33×10$^{-4}$ | −42.2 |
| ML | [18], Fig. 3 (X89) | na | 2.45×10$^{-4}$ | −9.0 | 4.15×10$^{-4}$ | −19.0 |
| FN | [19], Fig. 8 | Fig. 2 | 1.0×10$^{-4}$ | −28.5 | 2.0×10$^{-4}$ | −36.9 |

Table 4. Output data from orthodoxy test.

| Plot type | Figure here | $f_{\text{low}}^{\text{extr}}$ | $f_{\text{up}}^{\text{extr}}$ | Result |
|---|---|---|---|---|
| MG | Fig. 1 | 0.15 | 0.35 | PASS |
| ML | na | 0.20 | 0.34 | PASS |
| FN | Fig. 2 | 0.26 | 0.52 | INCONCLUSIVE |

It needs to be remembered that this test is an "engineering triage test", with somewhat arbitrary boundaries for the three categories of "pass", "fail", and "inconclusive". The "pass" and "fail" categories have been set so that outcomes in these categories are reasonably certain. The "inconclusive" category can therefore cover both situations that are "nearly normal" and and others that are "definitely not normal".

## 5. Summary and conclusions

This paper has set out in a concise form the theory behind the orthodoxy test, and has shown that, in addition to its current use with Millikan-Lauritsen plots and Fowler-Nordheim plots, it can also easily be applied to Murphy-Good plots.

It has been argued elsewhere [11] that MG plots provide a better methodology of FE current-voltage data analysis than do FN plots, because for ideal FE devices/systems they lead to the more precise extraction of information about formal emission area. The work in this paper confirms that, in addition, MG plots can be subject to the orthodoxy test that is the necessary preliminary to meaningful data analysis.

We have also reported the initial development of a prototype web tool that can carry out the orthodoxy test for all three types of $I(V)$ data plot. Further development of this web tool is in progress, in order to allow the extraction of characterization parameters from plots that pass the test. Our intention is that, in future work on carbon field emitters at Mu'tah University, Murphy-Good plots and the related form of orthodoxy test will be used. It is also our intention to develop a downloadable spreadsheet version of the web application.

## Appendix A: The Schottky constant

The "Schottky constant" is the modern equivalent of the numerical constant that appeared in eq. (6) of the 1914 paper [20] that first put the theory of the Schottky effect (see Wikipedia) onto a quantitative basis. Since the term "Schottky constant" may be unfamiliar to many people, relevant background information is presented here.

A classical point electron escaping from a surface to which a high negative field (of magnitude $F$) is applied experiences forces due to both its electrical image in the surface and the external electrostatic field. As a consequence, as compared with the classical potential energy (PE) barrier that would be seen by the electron when $F = 0$, the escaping electron experiences a classical PE barrier with a maximum height that has been reduced by an energy $\Delta_\mathrm{S}$ given by

$$\Delta_\mathrm{S} = c_\mathrm{S} F^{1/2}.$$

This is the well-known classical *Schottky effect*, in fact first suggested as an electron emission mechanism by J.J. Thomson in 1903 [21]. The parameter $c_\mathrm{S}$ is a universal constant that has now been called the *Schottky constant*.

In terms of the fundamental physical constants, $c_\mathrm{S}$ is given by [15]:

$$c_\mathrm{S} = (e^3/4\pi\varepsilon_0)^{1/2},$$

where $e$ is the elementary charge and $\varepsilon_0$ the vacuum electric permittivity. In the units now often used in field emission, $c_\mathrm{S}$ has the value 1.199985 eV (V/nm)$^{-1/2}$, and $c_\mathrm{S}^2$ has the value 1.438865 eV$^2$ (V/nm)$^{-1}$.

As already noted, the Schottky effect was first put on a quantitative basis in Schottky's 1914 paper [20]. If fields are measured in V/cm, as often done before the SI system was introduced, then the modern parameter $c_\mathrm{S}$ has the value 3.794686×10$^{-4}$ eV (V/cm)$^{-1/2}$. The corresponding numerical value, approximated as 3.8×10$^{-4}$, appears in eq. (6) of [20]. This is the origin of the name "Schottky constant" for $c_\mathrm{S}$. In one form or another, the physics of the Schottky constant has been in use for over 100 years, but only recently has $c_\mathrm{S}$ been explicitly given this name.

**Acknowledgments**

The authors would like to thank the Deanship of Academic Research at Mu'tah University for supporting this work through research project number 241/19/120. The research was also supported by the Ministry of Industry and Trade of the Czech Republic, MPO-TRIO project FV10618, and by the University of Surrey, UK.

**Mohammad M. Allaham (student)** was born in 1986 in Amman, Jordan. He is currently an MSc student in Mutah University, Al-Karak, Jordan and part of a research team led by Prof. Marwan S. Mousa. The team investigates the development of new experimental electron sources, along with a theoretically oriented team which analyses data obtained from the experiments.

**Richard G. Forbes (DSc)** studied at Trinity College, Cambridge, where he received a Tripos Prize, a Senior Scholarship and a Research Scholarship. After a first degree in Physics (Theoretical Option), he did PhD research in the Dept. of Metallurgy and Materials Science in Cambridge. He did postdoctoral work at Aston University, subsequently becoming a Lecturer in Physics. He moved to Surrey University in 1983, and retired as Reader in Applied Electrophysics. He continues to be active in research, as a Visiting Reader. Dr Forbes is well known internationally for work in high-electric-field nanoscience and in field electron emission. Descriptions of much of his work can be found on ResearchGate.

**Alexandr Knápek (PhD)** was born in 1983 in Brno, Czech Republic. He received his PhD from Brno University of Technology in Brno in 2013 where he also continued to lecture Physics to bachelor students till 2017. Currently, he works as a researcher at the Institute of Scientific Instruments of the Czech Academy of Sciences within the Electron-beam

Lithography Group. There, he deals mainly with experimental research of field emission emitters and other topics within the field of physical electronics.

**Marwan S. Mousa (Prof., PhD)** is affiliated to the Dept. of Physics, Mu'tah University, where he has been working since 1985. Prof. Marwan Mousa has authored and co-authored more than 150 national and international publications. His research interests focus mainly on field emission of electrons from composite electron sources. He has an active association with different societies and academies around the world, leaving his mark in the scientific community. Dr. Marwan Mousa became a full professor in Physics and Materials Technology in May 1996, and has received several awards for contributions to the scientific community. In October 2019, he was appointed Vice-President of Mu'tah University.